\documentclass{WileyMSP-template}
\usepackage{textcomp,gensymb,amsmath,xcolor, graphicx}
\usepackage{hyperref}
\usepackage{pdfpages}
\bibstyle{plain}

\newcommand{\txc}{}
\begin{document}


\title{Contact-Barrier Free, High Mobility, Dual-Gated Junctionless Transistor Using Tellurium Nanowire}

\maketitle


\author{Pushkar Dasika}
\author{Debadarshini Samantaray}
\author{Krishna Murali}
\author{Nithin Abraham}
\author{Kenji Watanabe}
\author{Takashi Taniguchi}
\author{N. Ravishankar}
\author{Kausik Majumdar}



\begin{affiliations}
Pushkar Dasika, Krishna Murali, Nithin Abraham\\
Department of Electrical Communication Engineering,\\
Indian Institute of Science, Bangalore 560012, India \\
Email Address:kausikm@iisc.ac.in \\

Debadarshini Samantaray, N. Ravishankar\\
Materials Research Center, \\
Indian Institute of Science, Bangalore 560012, India

Kenji Watanabe \\
Research Center for Functional Materials, National Institute for Materials Science,\\
1-1 Namiki, Tsukuba 305-044, Japan

Takashi Taniguchi \\
International Center for Materials Nanoarchitectonics, National Institute for Materials Science, \\
1-1 Namiki, Tsukuba, 305-044 Japan

\end{affiliations}



\begin{abstract}
Gate-all-around nanowire transistor, due to its extremely tight electrostatic control and vertical integration capability, is a highly promising candidate for sub-5 nm technology node. In particular, the junctionless nanowire transistors are highly scalable with reduced variability due to avoidance of steep source/drain junction formation by ion implantation. Here we demonstrate a dual-gated junctionless nanowire \emph{p}-type field effect transistor using tellurium nanowire as the channel. The dangling-bond-free surface due to the unique helical crystal structure of the nanowire, coupled with an integration of dangling-bond-free, high quality hBN gate dielectric, allows us to achieve a phonon-limited field effect hole mobility of $570\,\mathrm{cm^{2}/V\cdot s}$ at 270 K, which is well above state-of-the-art strained Si hole mobility. By lowering the temperature, the mobility increases to $1390\,\mathrm{cm^{2}/V\cdot s}$ and becomes primarily limited by Coulomb scattering. \txc{The combination of an electron affinity of $\sim$4 eV and a small bandgap of tellurium provides zero Schottky barrier height for hole injection at the metal-contact interface}, which is remarkable for reduction of contact resistance in a highly scaled transistor. Exploiting these properties, coupled with the dual-gated operation, we achieve a high drive current of $216\,\mathrm{\mu A/\mu m}$ while maintaining an on-off ratio in excess of $2\times10^4$. The findings have intriguing prospects for alternate channel material based next-generation electronics.
\end{abstract}


\section{Introduction}
The tight electrostatic control in multi-gated nanowire field effect transistor (FET) \cite{cui2003high,singh2006high,appenzeller2008toward,kuhn2012considerations} makes it one of the most promising candidates for further scaling of silicon technology in sub-5 nm technology nodes, and is being actively pursued by the semiconductor industry \cite{bohr2017cmos,Samsung_Plans_Mass}. The possible vertical integration of multiple nanowires enhances the drive current at a given footprint \cite{lee2016vertically,Samsung_Plans_Mass}.
In parallel, several high-mobility, non-silicon channel options are being actively pursued as well, particularly for the p-FET \cite{ikeda2012high,fang2007vertically,wang2003germanium}.

Three technological roadblocks must be overcome to achieve these high performance, ultra-scaled devices. First, the source-drain electrical isolation in a transistor is usually achieved by doping. The required spatial steepness of the doping profile is non-trivial to achieve in such ultra-short channel nanowire FETs. It also adds to the large variability, resulting in low yield. In this direction, junctionless FET has attracted a lot of attention for high scalability \cite{lee2009junctionless,gundapaneni2011enhanced} where one uses a narrow conducting channel, and the gate voltage is used to deplete the carriers from the channel to turn it off, thus avoiding source-drain junction formation through ion implantation. Second, since the nanowire must be very narrow to achieve the required gate control, the surface roughness scattering is enhanced \cite{jin2007modeling,ramayya2008electron}, significantly degrading the carrier mobility. A nanowire channel with dangling-bond-free surface will be ideally suited to reduce such mobility degradation. Third, as the device footprint goes down with transistor scaling, the effective contact area at the source and drain reduces. This increases the fraction of the contact resistance in the total transistor resistance under on condition, which is detrimental to the overall drive current, switching speed, current saturation, and output resistance \cite{majumdar2010effects}. One must use novel techniques to reduce contact resistivity.

In an attempt to address the above-mentioned issues, here we use a dual-gated tellurium (Te) nanowire junctionless transistor. The unique chiral nature of the nanowire crystal structure \cite{liang2009synthesis,qin2020raman} allows a high hole mobility due to dangling-bond-free surface, and its low electron affinity helps us to achieve zero Schottky barrier at the contact interface, reducing the contact resistance. We achieve a drive current of $216\,\mathrm{\mu A/\mu m}$ with an on-off ratio of $> 2\times 10^4$ and a high hole mobility of $570\,\mathrm{cm^{2}/V\cdot s}$.

Te is an element in the chalcogen group, and its crystal structure consists of covalently bonded one-dimensional chains of Te along the c-axis that are held together by weak van der Waals interactions [inset of Fig. \ref{fig:material_char}(a)]. Thus, the nanowire form of Te is predominantly bound by lower energy surfaces that are produced by only breaking the weak van der Waals bonds and not the strong covalent bonds, leading to dangling-bond-free surface of the nanowire. This unique anisotropic crystal structure of Te provides an opportunity to synthesize nanowires of Te down to single chain Te atoms \cite{medeiros2017single,qin2020raman}.

Tellurium, in its bulk form, has a bandgap of 0.35 eV and exhibits \emph{p}-type nature \cite{zhao2020evaporated} with high carrier
mobility \cite{zhu2017multivalency}. As the bulk Te is reduced to smaller dimensions, it starts exhibiting an indirect bandgap of about 0.6 eV \cite{roy2017manipulation,qin2020raman}. In addition to these electrical properties, Te also shows remarkable optical, thermoelectric, and piezoelectric properties \cite{lee2013high,pradhan2017ultra,qiu2019thermoelectric,lin2016tellurium}.

\section{Synthesis and fabrication}

Te nanowires are synthesized (see \textbf{Methods} for details) using $\mathrm{Na_{2}TeO_{3}}$
as a precursor in the water medium, where hydrazine hydrate is used as a reducing
agent, PVP as a capping agent, and Ammonia is used to maintain proper
pH. After the reaction and cool down to room temperature, the solution is sonicated
twelve times using hot water ($60^{\circ}\mathrm{C}$) at 8000 revolutions per minute and subsequently two times using ethanol.

\paragraph{Structural and microstructural characterization of Te nanowire:}
As synthesized Te nanowires have been characterized by X-ray diffraction (XRD) using Rigaku X-ray Diffractometer with Cu $\mathrm{K}\alpha$ source. Fig. \ref{fig:material_char}(a) represents the XRD pattern of the nanowire which corresponds to the rhombohedral phase
($\mathrm{P3_{1}21}$) of Te (JCPDS : 361452) with lattice parameters
of $\mathrm{a=b=4.45\,\mathring{A}}$, $\mathrm{c=5.9\,\mathring{A}}$).
Transmission electron micrograph (TEM) in Fig. \ref{fig:material_char}(b) shows the one-dimensional morphology
as well as the diameter distribution of the nanowires. The nanowires have an average diameter of 50 nm and a length
in tens of micrometers. The selected area electron diffraction (SAED) pattern
captured from one such nanowire is shown in Fig. \ref{fig:material_char}(c)
which suggests the single-crystalline nature of the nanowire. The diffraction
spots correspond to {[}$1\bar{1}0${]} zone axis pattern and can be matched
to ($002$) and ($110$) reflections of Te. The lattice fringes in high
resolution TEM image in Fig. \ref{fig:material_char}(d) confirms the growth direction
to be {[}001{]}. The high angle annular dark field (HAADF) STEM image of the nanowire is shown in Fig. \ref{fig:material_char}(e) where the atomic layers of Te are clearly visible.

The nanowires exhibit a strong Raman peak [Fig. \ref{fig:material_char}(f)] at $122.8\,\mathrm{cm^{-1}}$ corresponding to the $A_1$  mode of lattice vibration at the Brillouin zone center \cite{pine1971raman,du2017one}. The $E_{TO}$ zone center modes of vibration manifest as another strong Raman peak at $142.6\,\mathrm{cm^{-1}}$ and a weak shoulder around $103\,\mathrm{cm^{-1}}$ \cite{pine1971raman,liu2010rapid}. We assign the broad peak around $265\,\mathrm{cm^{-1}}$ to second order Raman peak corresponding to $A_1+E_{TO}$ two-phonon process \cite{song2008superlong}.

\paragraph{Device Fabrication:}
The step-by-step fabrication process of the junctionless FET is illustrated in Fig. \ref{fig:plasma_clean}(a). We drop cast the Te nanowires onto a degenerately
doped Si substrate coated with 285 nm thick, thermally grown, high quality $\mathrm{SiO_{2}}$. A low power Ar treatment is then performed
to remove any residual PVP coating from the nanowires. A thin layer
of high-quality hBN ($\sim$ 15 nm) is immediately transferred onto the nanowire
using a dry transfer technique \cite{Castellanos_Gomez_2014} to define the top gate dielectric. {The atomically smooth surface of dangling-bond-free hBN gate dielectric provides an ideal interface with the dangling-bond-free Te nanowire channel.}
Subsequently, electron beam lithography is performed to define the source, the drain, and the gate metallization areas. To remove any unintentional residue and surface oxidation \cite{lan2007dispersibility} from the source/drain region of the
nanowire, a second plasma cleaning step is performed, which improves the contact interface quality. Ni/Au is then deposited using DC magnetron sputtering, followed by lift-off in acetone to complete the device fabrication. More details of the fabrication are provided in \textbf{Methods}.

We find that an optimum plasma cleaning is a crucial step in the fabrication, which helps remove residual PVP capping layer and possible interfacial oxide, improving the interface quality and reducing the contact resistance. A careful optimization shows that Ar plasma cleaning for 20 s at a power level of 10 W and a pressure of 10 mbar results in the optimum device performance. The Figs. \ref{fig:plasma_clean}(b)-(c)
show the STEM images of two different nanowires, one without the plasma cleaning
and the other after plasma cleaning for 20 s, respectively. The plasma cleaned nanowire clearly shows superior surface quality. The electrical characteristics of a nanowire without plasma cleaning are shown in Fig. \ref{fig:plasma_clean}(d) at various temperatures from 7 to
200 K. The characteristics show very low current levels and fast current saturation.
Such an observation has also been made previously for devices using two-dimensional layer materials
\cite{ho2017high}. On the other hand, the nanowires with plasma cleaning show a 1000-fold enhancement in the current drive at similar external bias, indicating superior contact quality as shown in \ref{fig:plasma_clean}(e)

\section{Results and discussion}
We measured several devices with a typical diameter of the nanowires chosen between 30 to 50 nm. \txc{The nanowires, with a low bandgap} \cite{haeffler1996electron}, exhibit strong p-type conductivity, even without any gating. This allows us to operate the fabricated transistor as a dual -gated junctionless nanowire FET, as schematically illustrated in Fig. \ref{fig:dual_gate}(a). The plan view SEM image of a typical nanowire FET is shown in Fig. \ref{fig:dual_gate}(b). The cross-section of the device along the white dashed line is provided in \textbf{Supporting Figure 1}. The transistor is normally in the on state and requires a positive gate voltage to deplete the holes from the channel (depletion mode). The top gate using the hBN dielectric controls the channel conductivity, while the global back gate can control both the channel portion, as well as the underlap regions and the metal contact region. This allows us to avoid the need for source/drain doping, and the device works as a junctionless transistor.

The modulation of the drain current ($I_D$) as a function of both the top ($V_{TG}$) and the bottom ($V_{BG}$) gate voltages is shown as a two-dimensional color plot in Fig. \ref{fig:dual_gate}(c), at $V_D=-1.2$ V. A nanowire FET, owing to the improved electrostatic control due to their cylindrical geometry, exhibits significantly improved on-off ratio compared to transistors with planar structure using tellurium nanosheet of similar thickness \cite{wang2018field}. The three line cuts along the dashed lines (i-iii) in Fig. \ref{fig:dual_gate}(c) are shown in Fig. \ref{fig:dual_gate}(d). The dual-gate structure along the off-diagonal dashed line (iii) achieves superior gate control compared with stand-alone $V_{TG}$ [along (i)] or $V_{BG}$ [along (ii)]. Since the transistor is operated in depletion mode, the application of a negative $V_{TG}$ and $V_{BG}$ does not improve the drive current significantly, as can be seen from the
characteristics. However, a positive gate voltage efficiently modulates $I_D$ by depleting the holes from the channel. While the obtained on-off ratio is $\sim 10^2$ using either the top or the bottom gate separately, we achieve an on-off ratio over $2\times10^4$ when the two gates are simultaneously used. The results of a coupled electrostatics and carrier transport simulation from a junctionless transistor are summarized in \textbf{Supporting Figure 2}.


The output characteristics of the device are shown in Fig. \ref{fig:dual_gate}(e-f), which are obtained by sweeping $V_{TG}$ while keeping $V_{BG}$
at $30$ and $-30$ V, respectively.
Since the global back gate can modulate the effective barrier to hole
injection, at $\mathrm{V_{BG}=30\,V}$, the barrier to hole injection
is high, and hence the output characteristics show a predominantly
Schottky behaviour. However, at $\mathrm{V_{BG}=-30\,V}$, the barrier to
hole injection is removed, and the underlap regions are electrostatically p-doped, leading to output characteristics tending towards current saturation behavior. Thus, by changing the biasing configuration, we are able to convert the device operation from a Schottky barrier FET (SBFET) to a MOSFET. {Similar architecture has previously been employed with Silicon nanowire transistors \cite{koo2005enhanced,colli2009top} wherein a combination of a top gate and a back gate has been utilized to improve the on-off ratio.} In the MOSFET mode, we achieve a high drive current of $216\,\mathrm{\mu A/\mu m}$ (calculated using $\frac{I_D}{2\pi r}$ where $r$ is the radius of the nanowire) from the nanowire FET. Given the relaxed geometry used in this work, this is very promising towards high performance scaled transistor.

A qualitative band diagram of metal (Ni) - Te junction is shown in
Fig. \ref{fig:low_temp}(a) at equilibrium. \txc{The electron affinity of Te
$\chi_{Te}\approx 4\,\mathrm{eV}$ \cite{wang2018field}} and
Ni has a work function $\phi_{m}\approx 5\,\mathrm{eV}$. The p-type conductivity suggests that
the Fermi level in Te is near the valence band, and thus we can
see that $\phi_{m}\gg\phi_{s}$ where $\phi_{s}$ is Te work
function. This allows the holes to move freely across the contact interface without any Schottky barrier and is ideal for an excellent ohmic contact \cite{pierret1996semiconductor}.

To analyze such contact barrier-free operation in the MOSFET mode, we measure the temperature-dependent device characteristics keeping the bottom gate floating, and varying $V_{TG}$ and $V_D$. The transfer characteristics of the device from 5 to 270 K are shown in Fig. \ref{fig:low_temp}(b). It is striking that under on condition ($V_{TG}<0$), $I_D$ is a very weak function of temperature, even down to temperatures as low as 5 K, and $I_D$ increases with a reduction in temperature (due to a subsequent increase in hole mobility, as discussed later) - clearly suggesting barrier-free hole injection from the contact.

We further verify this using modified Richardson equation for nanowire \cite{o2008thermionic} in Fig. \ref{fig:low_temp}(c), wherein the inset, we plot $\frac{I_D}{T}$ in the log scale as a function of $\frac{1}{T}$ at various $V_{TG}$. The effective barrier height, as extracted from the slope of the linear fits, is shown in Fig. \ref{fig:low_temp}(c). The positive slope (and hence the `negative' effective barrier) under on state is due to a reduction in $I_D$ with $T$, representing a negligible barrier at the contact interface. This is independent of $V_{TG}$ as the top gate only controls the channel portion. At larger positive $V_{TG}$, the extracted effective barrier increases, which corresponds to the $V_{TG}$ induced potential barrier created between the source underlap and the channel, as the channel turns off. The output characteristics of the device shown in Fig. \ref{fig:low_temp}(d) further supports the negligible Schottky barrier even at $T=5\,K$. The characteristics show strong current saturation, particularly at relatively low $|V_{TG}|$, and $I_D$ varies approximately in a quadratic manner with $V_{TG}$.

Using the temperature dependent transfer characteristics, we extract the field effect mobility ($\mu$) of the nanowire FET as \cite{wang2003germanium,sze2006physics}
\begin{equation}
\mu=\frac{dI_{D}}{dV_{TG}}\frac{L_{ch}}{C_{ox}}\frac{1}{V^\prime_{D}}
\end{equation}
Here $C_{ox}$ is the gate capacitance per unit length of the device,
$L_{ch}$ is the channel length of the device, and $V'_{D}$ is the effective drain bias after correcting for the series resistance.
Since the gate does not entirely cover the device, there are significant
underlap regions which contribute to the series resistance. This effect
can be seen in output characteristics shown in Fig. \ref{fig:low_temp}(d)
where at high $|V_{TG}|$, the rate of increase of current with
$V_{TG}$ decreases, indicating that $I_D$ is
limited by the series resistance. The total series resistance is estimated as
\begin{equation}
R_{s}=R_{D}\frac{L_{under}}{L_{wire}}
\end{equation}
where $R_D$ is the output resistance in the linear regime with $V_{TG}\approx 0$, $L_{wire}$ is the total length of the nanowire and $L_{under}$ ($=L_{wire}-L_{ch}$) is the
total length of the underlap regions. The corrected drain bias then becomes
\begin{equation}
V^\prime_{D}=V_{D}-I_{D}R_{s}
\end{equation}
where $V_{DS}$ is the applied drain bias.

The channel being cylindrical in nature, $C_{ox}$ is very sensitive to device geometry and can vary up to 10-fold if
the correct device structure is not taken into account \cite{khanal2007gate,wunnicke2006gate,fuhrer2013measurement}. The cross sectional geometry of the device and the corresponding scanning electron micrograph are shown in Fig. \ref{fig:mob}(a-b).
To accurately capture the gate capacitance, Laplace equation is solved
for this device geometry
\begin{equation}
\frac{\partial^{2}\phi}{\partial x^{2}}+\frac{\partial^{2}\phi}{\partial y^{2}}=0
\end{equation}
using finite element method (FEM) as implemented in FreeFEM++ software \cite{MR3043640}.
The nanowire is assumed to be conducting and hence is equipotential
with Dirichlet boundary conditions $\phi=V$. The top surface of hBN
is assumed to have fixed potential at $\phi=0$\@. Since the bottom gate in the measurement
is left floating, the electric field at the bottom of 285 nm $\mathrm{SiO_{2}}$
is assumed to be zero. The total charge ($Q$)
on the nanowire is calculated by integrating the normal component of the
electric field on the nanowire. The gate capacitance is then obtained from $C=\frac{Q}{V}$.
The electrostatic potential contour lines and the corresponding electric field lines
as obtained from the 2D FEM simulation is shown in
Fig. \ref{fig:mob}(d). {Due to the relatively larger diameter of the nanowires used in this work, we have neglected the quantum capacitance of the channel \cite{fickenscher2013optical, pemasiri2015quantum, lee2012gate} arising from the finite density of states. This quantum capacitance comes in series with the gate oxide capacitance calculated above. This results in a slight overestimation of the gate capacitance, hence underestimating the extracted carrier mobility. Thus the extracted hole mobility values discussed next represent the lower bounds of the true mobility.}

The extracted $\mu$ is shown in Fig. \ref{fig:mob}(c) as a function of $V_{TG}$ at different $T$. $\mu$ has a peak ($\mu_{peak}$) near the threshold voltage, where it shows strong temperature dependence. On the other hand, at higher negative $V_{TG}$, it drops due to large gate field \cite{takagi1994universality}, and the strong temperature dependence vanishes. The nanowire FET exhibits $\mu_{peak}=570\,\mathrm{cm^{2}/V\cdot s}$ at 270 K, which is twice of (110) Si hole mobility, and also beats the state-of-the-art uniaxially compressed (110) strained Si hole mobility \cite{kuhn2012considerations}. $\mu_{peak}$ increases monotonously with a decrease in the temperature initially, and gradually saturates to a value of $1390\,\mathrm{cm^{2}/V\cdot s}$ at low temperature [Fig. \ref{fig:mob}(e)]. For $\mathrm{T>180\,K}$, $\mu_{peak}$ can be fitted to a power law:
\begin{equation}
\mu_{peak}\propto T^{-\gamma}
\end{equation}
with $\gamma=1.47$. This indicates that for $\mathrm{T>180\,K}$,
$\mu_{peak}$ is limited by phonon scattering, while at lower temperatures,
it is limited by Coulomb scattering \cite{ong2013mobility,radisavljevic2013mobility}.
For solution synthesized Te nanoflakes, $\gamma$ has been found to
be 1.03 \cite{amani2018solution} while in Te thin films grown by molecular
beam epitaxy, $\gamma$ is found to vary between 0.75 to nearly zero
\cite{zhou2018high}.

{A comparison of performance of the presented Te nanowire transistor with other reported Te-based transistors \cite{wang2018field,liang2009synthesis,zhou2018high,amani2018solution,tong2020stable,qin2020raman} is shown in Fig. \ref{fig:mob}(f-g) in terms of drive current and on-off ratio. To make a fair comparison by accommodating varied device dimensions and biasing conditions in reported data from the literature, we scale the drive current ($I_{ON}$) by the drain field and the circumference of the nanowire (channel width in case of 2D flakes) as follows:
\begin{equation}I_{scaled} = \frac{I_{ON} L_{wire}}{2\pi r V_{DS}}
\end{equation}
Previous reports suggest that Te exhibits high mobility for thicker films (and wider nanowires); however, it becomes challenging to turn the channel off at such channel thickness. On the other hand, for thin channels, where the on-off ratio is improved, the carrier mobility degrades significantly, and Te loses its attractiveness as a high-mobility channel. Fig. \ref{fig:mob}(f) shows the on-off ratio achieved in various Te-based transistor structures reported earlier as a function of the channel thickness (or diameter), suggesting a strong degradation of the on-off ratio for thicker channels. From the same figure, we observe that the advantage of the improved electrostatics \txc{due to dual-gated nanowire structure in the present work is remarkable, allowing us to achieve an on-off ratio of $2\times10^4$ ($@V_{ds}=-1.2$ V) at a relatively larger nanowire diameter. However, the device's subthreshold swing is still relatively large, and there is scope for further optimization on the gate electrostatics and nanowire-dielectric interface.}  In Fig. \ref{fig:mob}(g), we compare the performance of different reports in the on-off ratio and drive current space. We note from the figure that the present work provides comparable performance with the best reports in terms of drive current, while maintaining a superior on-off ratio.


In summary, we presented the synthesis, fabrication, and electrical characterization
of high-performance dual-gated p-type Te nanowire junctionless transistors. The unique combination of contact-barrier-free hole injection, high hole mobility ($570$ and $1390\,\mathrm{cm^{2}/V\cdot s}$ at 270 and 5 K, respectively), {dangling-bond-free, clean interface between the nanowire channel and hBN dielectric,} and dual-gated operation allows us to achieve an on-off ratio in excess of $2\times10^{4}$ with a high on-current of $216\,\mu A/\mu m$ even at a relatively small drain field, which is superior to existing p-type nanowire FETs. Te nanowires are thus a promising candidate for next-generation ultra-scaled transistors, as well as for other nanowire-based electronic applications, including highly scaled memory, biosensing, fast infrared sensing, and spectroscopy.

\section*{Methods}

\subsection*{Chemical synthesis of Te nanowires}
The Te nanowires are synthesized \cite{roy2017manipulation} using $\mathrm{Na_{2}TeO_{3}}$
as a precursor in the water medium, where hydrazine hydrate is used as a reducing
agent, Polyvinylpyrrolidone( PVP) as a capping agent, and Ammonia is used to maintain proper
pH. In a typical experiment, 1 g of PVP (average. M.W.= 58,000)
is dissolved in 20 ml of DI water and 92 mg $\mathrm{Na_{2}TeO_{3}}$
is dissolved in 15 ml DI water separately. Both solutions are mixed
at room temperature. Into the above solution, 1.5 ml of Hydrazine hydrate
and 3.3 ml of 25\% aqueous ammonia solution are added drop-wise with
a moderate stirring. The transparent solution is then transferred
to a 50 ml capacity teflon container. Then the hydrothermal reaction
is allowed to take place for 4 hours at $180\mathrm{^{\circ}C}$. After the
reaction vessel cools down to room temperature, the solution is cleaned
twelve times using hot water ($60^{\mathrm{\circ}}\mathrm{C}$) at 8000 revolutions per minute
and subsequently two times using ethanol.
\subsection*{Material and structural characterization}
X-ray diffraction measurement has been carried out using Rigaku X-ray
Diffractometer with Cu $\mathrm{K\alpha}$ source. Microstructural
characterization has been done using FEI Tecnai T20 S-Twin (200 kV)
and FEI-Titan G2 60-300 microscope operated at 300 kV. To get a cross-sectional view of the device, the desired pattern has been made using
FEI Helios G4 UX FIB instrument with Ga ion beam source operated at
30 kV (current $\sim 1.6$ nA) after depositing the device
with Pt ($\sim 3$ micron thick).
\subsection*{Device fabrication and characterization}
The solution synthesized nanowires were first drop cast on $\mathrm{p^{++}}$
Si substrate with 285 nm $\mathrm{SiO_{2}}$ grown on top. The substrate
is the Argon plasma cleaned at a power of 10 W and a pressure of 10 mbar for 20 s in PlasmaLab
system 100 ICP 380 from Oxford instruments. A thin layer of hBN is
then transferred using a dry transfer method. The substrate is then
spin-coated with PMMA C3 and baked on a hot plate at $180^\circ$ C for 2 minutes. e-Beam
lithography is then performed to define the source, drain, and gate contacts.
Patterns are then developed in MIBK:IPA solution in 1:3 ratio. Substrate
is then immediately plasma cleaned once again using the same parameters.
Ni/Au (20 nm/40 nm) is then deposited on the substrate using DC magnetron
sputtering. Finally, lift-off is performed by dipping the substrate
in acetone for 10 minutes, followed by drying.

Room temperature device characterization was done in ambient conditions
using a B1500 semiconductor parameter analyzer. Low-temperature characterization
was performed by loading the substrate in Montana cryostation using
a Keithley 4200A semiconductor parameter analyzer.

\medskip
\textbf{Acknowledgements} \par 
The authors acknowledge the electron microscopy facilities at the Advanced Facility for Microscopy and Microanalysis, IISc. K. M. acknowledges the support a grant from Indian Space Research Organization (ISRO), a grant from MHRD under STARS, grants under Ramanujan Fellowship and Nano Mission from the Department of Science and Technology (DST), Government of India, and support from MHRD, MeitY and DST Nano Mission through NNetRA. K.W. and T.T. acknowledge support from the Elemental Strategy Initiative conducted by the MEXT, Japan, Grant Number JPMXP0112101001, JSPS KAKENHI Grant Numbers JP20H00354 and the CREST(JPMJCR15F3), JST.
\medskip

%


\bibliography{references}
\bibliographystyle{MSP}


\begin{center}
\begin{figure}
\begin{centering}
\includegraphics[scale=0.5]{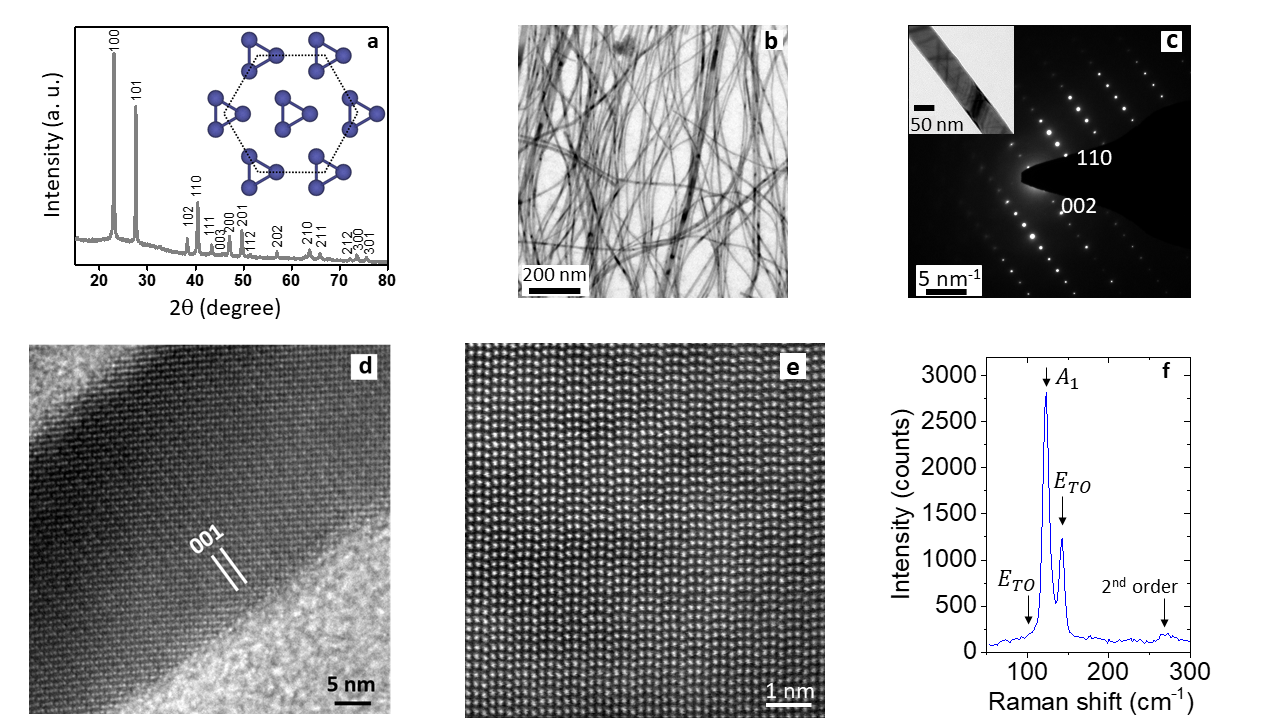}
\par\end{centering}
\caption{(a) X-ray diffraction pattern of Te
nanowires, which correspond to $P3_{1}21$ rhombohedral phase of Te. Inset: Crystal structure of hexagonal Te chains along the $c$ axis.
Solid lines indicate covalent bonding, while each chain is bonded
to other by van der Waals bonds.
(b) TEM image of the nanowires. (c) SAED from Te nanowire (bright-field image shown in the inset) showing single-crystalline nature
(corresponding to {[}$1\bar{1}0${]} zone axis). (d) HRTEM image of the Te nanowire
where the lattice fringes correspond to (001) plane of Te. (e) High-resolution HAADF-STEM image of Te nanowires showing the atomic layers.
(f) Raman spectroscopy of Te nanowire showing first order and second order peaks. \label{fig:material_char}}
\end{figure}
\par\end{center}

\begin{center}
\begin{figure}
\begin{centering}
\includegraphics[scale=0.5]{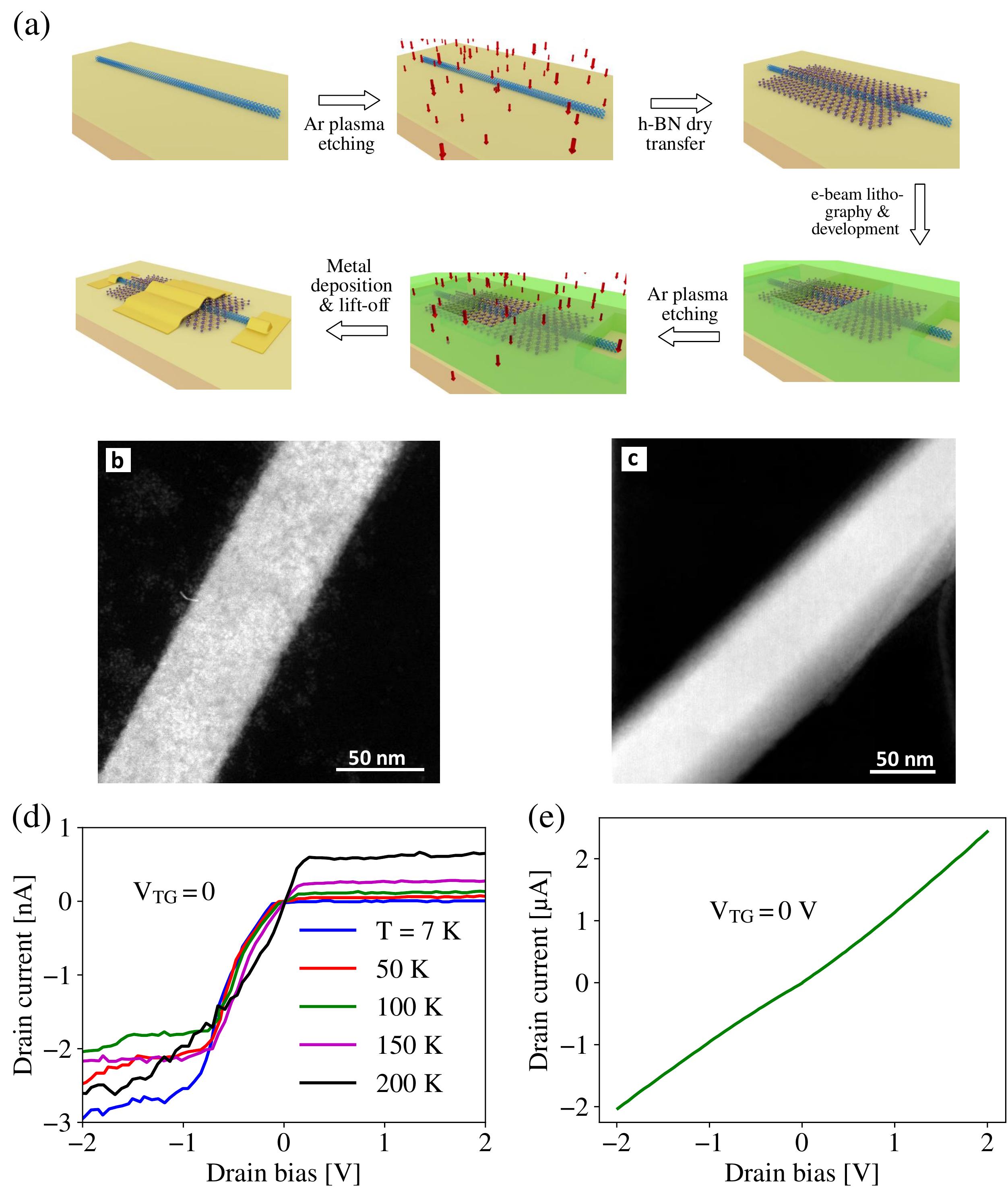}
\par\end{centering}
\caption{(a) Step-by-step process flow of Te nanowire junctionless FET fabrication. (b,c): HAADF-STEM image of Te nanowire (b) without plasma cleaning, and (c) with plasma cleaning for a duration of 20 sec. (d) Temperature dependent current-voltage characteristics of a nanowire device without plasma cleaning with low current levels and abrupt current saturation. (e) Characteristics of the device at room temperature  with plasma cleaning showing higher current levels and ohmic behaviour. \label{fig:plasma_clean}}
\end{figure}
\par\end{center}

\begin{center}
\begin{figure}
\begin{centering}
\includegraphics[scale=0.5]{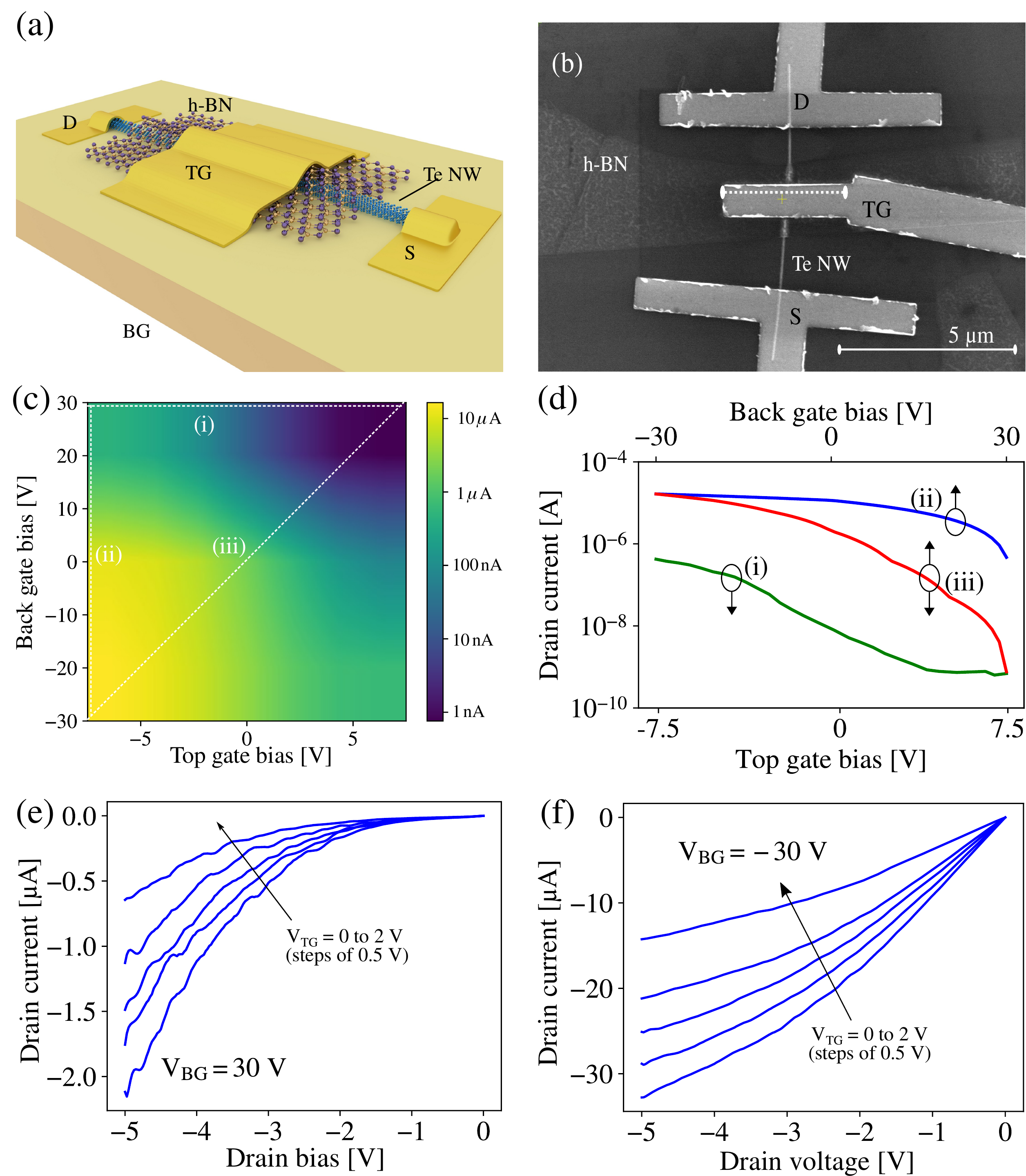}
\par\end{centering}
\caption{(a) Schematic representation of the dual-gated nanowire junctionless FET. (b) Scanning electron micrograph of the device. Scale bar is $5\mu$m. (c) Two-dimensional color plot of the drain current magnitude as the top gate $V_{TG}$ and back gate $V_{BG}$ voltages are varied at $V_{D}=-1.2\,V$. The bottom left corner
of the plot shows the on-state the device, while the upper right corner
shows the off-state. (d) The magnitude of the drain current of the device along the dashed lines shown in (c). Individual gates [top gate for (i) and bottom gate for (ii)] do not provide an on-off ratio of more than $10^2$, but dual-gating [dashed line along (iii)] shows an on-off ratio of $> 2\times10^4$. (e) Schottky-like output characteristics of the FET with varying $V_{TG}$ at a fixed positive back gate bias of $30$ V. (f) Output characteristics of the device at a fixed negative $V_{BG}$, measured at 295 K, showing MOSFET-like characteristics. \label{fig:dual_gate}}
\end{figure}
\par\end{center}

\begin{center}
\begin{figure}
\begin{centering}
\includegraphics[scale=0.5]{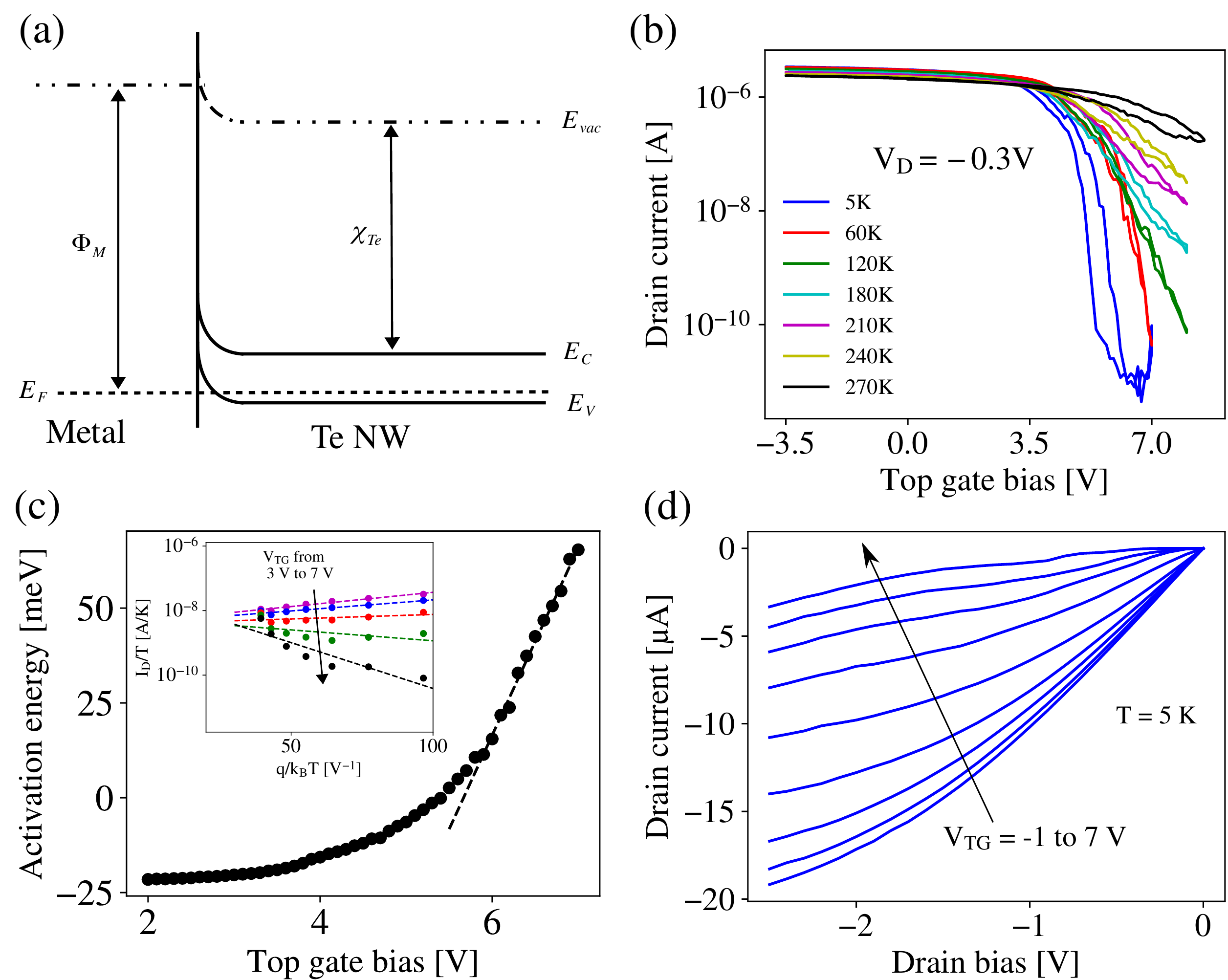}
\par\end{centering}
\caption{(a) Equilibrium band diagram of the device showing zero barrier to hole injection at the metal contact interface. (b) Transfer
characteristics of the device at various temperatures (showing both forward and reverse sweeps), keeping the back gate floating. Off current and subthreshold swing reduce steeply with decreasing temperature, while the on current increases. {The onset of ambipolar characteristics is observed at $T = 5$ K (blue trace).} (c) Extracted activation energy for hole injection in the device as obtained from the slope of the Richardson plot. Inset: Richardson plot depicting the dependence of drain current on the temperature at various $V_{TG}$ and $V_D = -0.6$ V. The negative extracted values at lower $V_{TG}$ correspond to the positive slope of the Richardson plot, which is due to a increase in $I_D$ with a reduction in temperature, indicating non-thermionic carrier injection at the contact interface. (d) Output characteristics of the device at $T = 5$ K. \label{fig:low_temp}}
\end{figure}
\par\end{center}

\begin{center}
\begin{figure}
\begin{centering}
\includegraphics[scale=0.43]{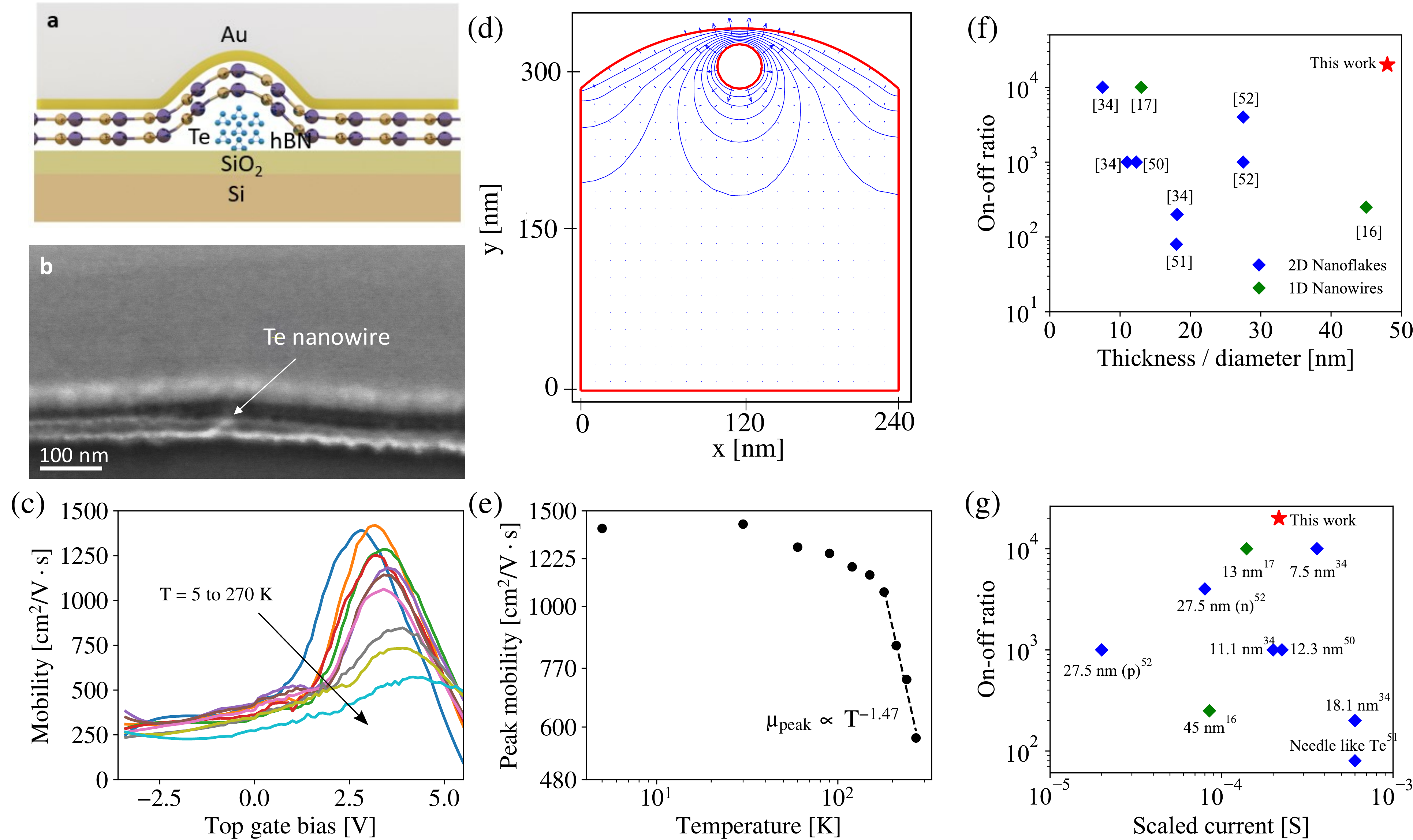}
\par\end{centering}
\caption{(a) Schematic cross-section of the nanowire device. (b) Scanning electron micrograph of the device cross-section when cut along the white dashed line in Fig. \ref{fig:dual_gate}(b) and corresponds to the schematic in (a). (c) Hole mobility \emph{versus} $V_{TG}$ at different temperatures, as extracted from the characteristics in Fig. \ref{fig:low_temp}(b). (d) Electrostatic potential contour lines and electric field lines of Te nanowire device, as obtained from 2D FEM simulation. The simulation is run on the device cross-section.  (e) Peak hole mobility plotted as a function of temperature. Mobility decreases from $1390\,\mathrm{cm^{2}/V\cdot s}$ at 5 K to $570\,\mathrm{cm^{2}/V\cdot s}$ at 270 K. The dashed line shows a fit of the peak mobility as $T^{-1.47}$ indicating phonon-limited mobility character. (f) Comparison of performance of various Te based field effect transistors in terms of on-off ratio and drive channel thickness/diameter.  (g) Comparison of performance of various Te based field effect transistors in terms of on-off ratio and drive current (scaled by drain field and nanowire circumference or channel width). The blue and red symbols indicate existing reports on 2D Te channel and 1D nanowire FETs, respectively, while the red star denotes the present work. \label{fig:mob}}
\end{figure}
\par\end{center}








\AtEndDocument{\includepdf[pages={2-3}]{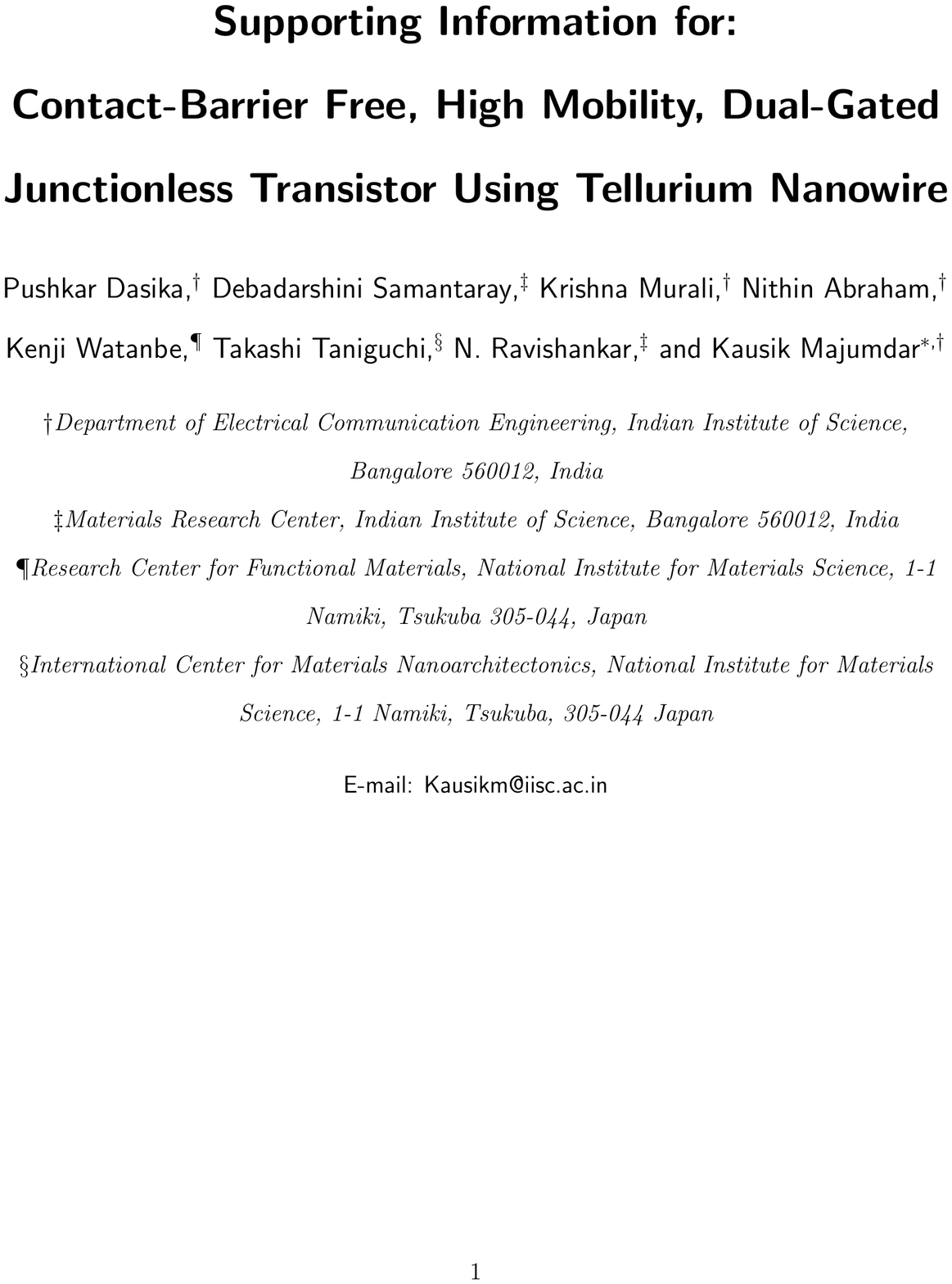}}
\end{document}